\documentclass[twocolumn]{aastex6}

\usepackage{amsmath}

\begin{document}

\title{Return of the King: Time-Series Photometry of FO Aquarii's Initial Recovery from its Unprecedented 2016 Low State}

\shorttitle{Photometry of FO Aqr's 2016 Low State} 
\shortauthors{Littlefield et al.}

\author{
Colin Littlefield,\altaffilmark{1}
Peter Garnavich,\altaffilmark{1}
Mark R. Kennedy,\altaffilmark{1,2}
Erin Aadland,\altaffilmark{1,3}
Donald M. Terndrup,\altaffilmark{4}
Grace V. Calhoun,\altaffilmark{4}
Paul Callanan,\altaffilmark{2}
Lyu Abe,\altaffilmark{5}
Philippe Bendjoya,\altaffilmark{5}
Jean-Pierre Rivet,\altaffilmark{5}
David Vernet,\altaffilmark{5}
Maxime Devog\`{e}le,\altaffilmark{5,6}
Benjamin Shappee,\altaffilmark{7}
Thomas Holoien,\altaffilmark{4}
Te\'{o}filo Arranz Heras,\altaffilmark{8}
Michel Bonnardeau,\altaffilmark{9}
Michael Cook,\altaffilmark{10}
Daniel Coulter,\altaffilmark{11}
Andr\'{e} Deback\`{e}re,\altaffilmark{12}
Shawn Dvorak,\altaffilmark{13}
James R. Foster,\altaffilmark{14}
William Goff,\altaffilmark{15}
Franz-Josef Hambsch,\altaffilmark{16, 17}
Barbara Harris,\altaffilmark{18}
Gordon Myers,\altaffilmark{19}
Peter Nelson,\altaffilmark{20}
Velimir Popov,\altaffilmark{21, 22}
Rob Solomon,\altaffilmark{23}
William L. Stein,\altaffilmark{24}
Geoff Stone,\altaffilmark{25}
Brad Vietje\altaffilmark{26}
}

\altaffiltext{1}{Department of Physics, University of Notre Dame, Notre Dame, IN, USA}
\altaffiltext{2}{Department of Physics, University College Cork, Cork, Ireland}
\altaffiltext{3}{Department of Physics and Astronomy, Minnesota State University Moorhead, Moorhead, MN, USA}
\altaffiltext{4}{Department of Astronomy, The Ohio State University, Columbus, Ohio, USA}
\altaffiltext{5}{Universit\'{e} C\^{o}te d'Azur, OCA, CNRS, Laboratoire Lagrange, Nice, France}
\altaffiltext{6}{Universit\'{e} de Li\`{e}ge, Space sciences, Technologies and Astrophysics Research (STAR) Institute, All\'{e}e du 6 Ao\^{u}t 19c, Sart Tilman, 4000 Li\`{e}ge, Belgium}
\altaffiltext{7}{Carnegie Observatories, 813 Santa Barbara Street, Pasadena, CA 91101, USA}
\altaffiltext{8}{Observatorio Las Pegueras de Navas de Oro (Segovia), Spain}
\altaffiltext{9}{MBCAA Observatory, Le Pavillon, 38930 Lalley, France}
\altaffiltext{10}{AAVSO, Newcastle Observatory, Newcastle, Ontario, Canada}
\altaffiltext{11}{Department of Physics and Astronomy, Michigan State University, East Lansing, MI, USA}
\altaffiltext{12}{LCOGT, Monistrol sur Loire, France}
\altaffiltext{13}{AAVSO, Rolling Hill Observatory, Lake County, Florida, USA}
\altaffiltext{14}{AAVSO/ARAS}
\altaffiltext{15}{AAVSO, 13508 Monitor Lane, Sutter Creek, CA, USA}
\altaffiltext{16}{AAVSO / Vereniging Voor Sterrenkunde (VVS), Brugge, Belgium}
\altaffiltext{17}{Bundesdeutsche Arbeitsgemeinschaft f\"{u}r Ver\"{a}nderliche Sterne e.V. (BAV), Berlin, Germany}
\altaffiltext{18}{AAVSO, New Smyrna Beach, FL}
\altaffiltext{19}{AAVSO, 5 Inverness Way, Hillsborough, CA, USA}
\altaffiltext{20}{AAVSO, Ellinbank Observatory, Australia}
\altaffiltext{21}{Department of Physics, Shumen University, Bulgaria}
\altaffiltext{22}{IRIDA Observatory, NAO Rozhen, Bulgaria}
\altaffiltext{23}{AAVSO, Perth, WA, Australia}
\altaffiltext{24}{AAVSO, 6025 Calle Paraiso, Las Cruces, NM, USA}
\altaffiltext{25}{AAVSO, Sierra Remote Observatories, Auberry, CA, USA}
\altaffiltext{26}{AAVSO:  Northeast Kingdom Astronomy Foundation, Peacham, VT, USA}

\begin{abstract}

In 2016 May, the intermediate polar FO~Aqr was detected in a low state for the first time in its observational history. We report time-resolved photometry of the system during its initial recovery from this faint state. Our data, which includes high-speed photometry with cadences of just 2 sec, shows the existence of very strong periodicities at 22.5 min and 11.26 min, equivalent to the spin-orbit beat frequency and twice its value, respectively. A pulse at the spin frequency is also present but at a much lower amplitude than is normally observed in the bright state. By comparing our power spectra with theoretical models, we infer that a substantial amount of accretion was stream-fed during our observations, in contrast to the disk-fed accretion that dominates the bright state. In addition, we find that FO~Aqr's rate of recovery has been unusually slow in comparison to rates of recovery seen in other magnetic cataclysmic variables, with an $e$-folding time of 115$\pm7$ days. The recovery also shows irregular variations in the median brightness of as much as 0.2~mag over a 10-day span. Finally, we show that the arrival times of the spin pulses are dependent upon the system's overall brightness. 

\end{abstract}


\section{Introduction} 

An intermediate polar (IP) is an interacting binary system featuring a low-mass donor star which overfills its Roche lobe, transferring mass to a magnetic white dwarf (WD) \citep[for a review, see ][]{patterson94}. As such, it is a subset of the cataclysmic variable stars (CVs), but the magnetism of the WD results in a number of characteristics which differentiate IPs from other CVs. After the accretion flow leaves the inner Lagrangian (L1) point, it follows a ballistic trajectory until it either (1) circularizes into an accretion disk whose inner region is truncated by the WD's magnetic field or (2) directly impacts the WD's magnetosphere \citep{hameury86}. In the former scenario (disk-fed accretion), the WD's magnetic field captures plasma from the inner accretion disk, while in the latter (stream-fed accretion), the plasma is captured at some point along its ballistic trajectory. In some cases, disk-fed and stream-fed accretion can occur simultaneously if part of the accretion stream overflows the disk following the stream-disk interaction.

Regardless of the mode of accretion, the plasma will begin to travel along the WD's magnetic field lines when the local magnetic pressure exerted by the WD exceeds the ram pressure of the accretion flow. During its journey along the field lines, the plasma travels out of the binary orbital plane, creating a three-dimensional structure known as an accretion curtain \citep{rosen88}. The material within the curtain finally impacts the WD near one of its magnetic poles and is shocked to X-ray-emitting temperatures.

\begin{figure}

\epsscale{1.2}
\plotone{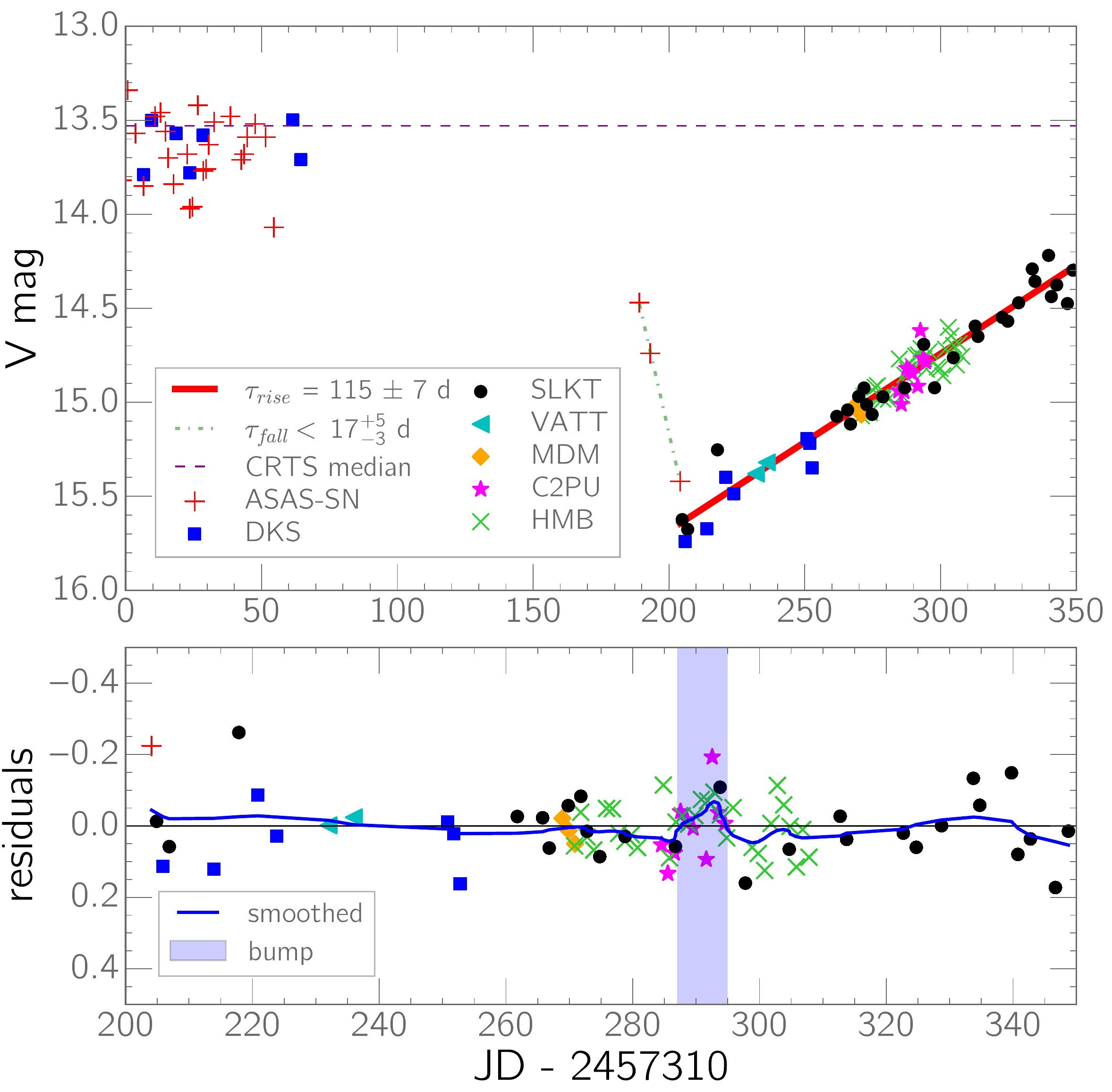}
\caption{The long-term light curve for FO Aqr, including the linear fit to the recovery (see Sec.~\ref{LCsec}). We also plot a line showing the slowest-possible single-sloped decline consistent with the three ASAS-SN observations between JD 189-204. The large gap results from solar conjunction. The `DKS' and `HMB' data are from co-authors Dvorak and Hambsch, respectively, and the other data are described in the text. In an effort to reduce contamination of the long-term light curve by short-term variations, the recovery data mostly show the median magnitude of the system during an extended time series; thus, ASAS-SN observations of the recovery are not plotted. The bottom panel shows residuals from the linear recovery model and uses Gaussian smoothing (blue line) to emphasize a bump in the light curve near JD 290 (shaded region). \label{longterm}}

\end{figure}

IPs show a complex range of periodicities in their optical light curves because the WD's spin frequency ($\omega$) is not synchronized to the orbital frequency ($\Omega$) of the system \citep{warner86}. If accretion is predominantly stream-fed, \citet{fw99} predict that the dominant frequency at optical wavelengths will be either the spin-orbit beat frequency ($\omega-\Omega$) or its first harmonic ($2\omega - 2\Omega$) if both poles are accreting and contributing equally to the light curve. The beat frequency is the rate at which the WD completes a full rotation within the binary rest frame. It is therefore the frequency at which the WD's magnetic field lines (rotating at $\omega$) will interact with stationary structures in the rest frame (rotating at $\Omega$), such as the accretion stream. If, however, the accretion is disk-fed, the energy released by accretion is independent of the orbital phase of the secondary, and optical variations will be seen at $\omega$ \citep{fw99}.

\begin{figure*}[ht!]

\centering
\epsscale{1.1}
\plotone{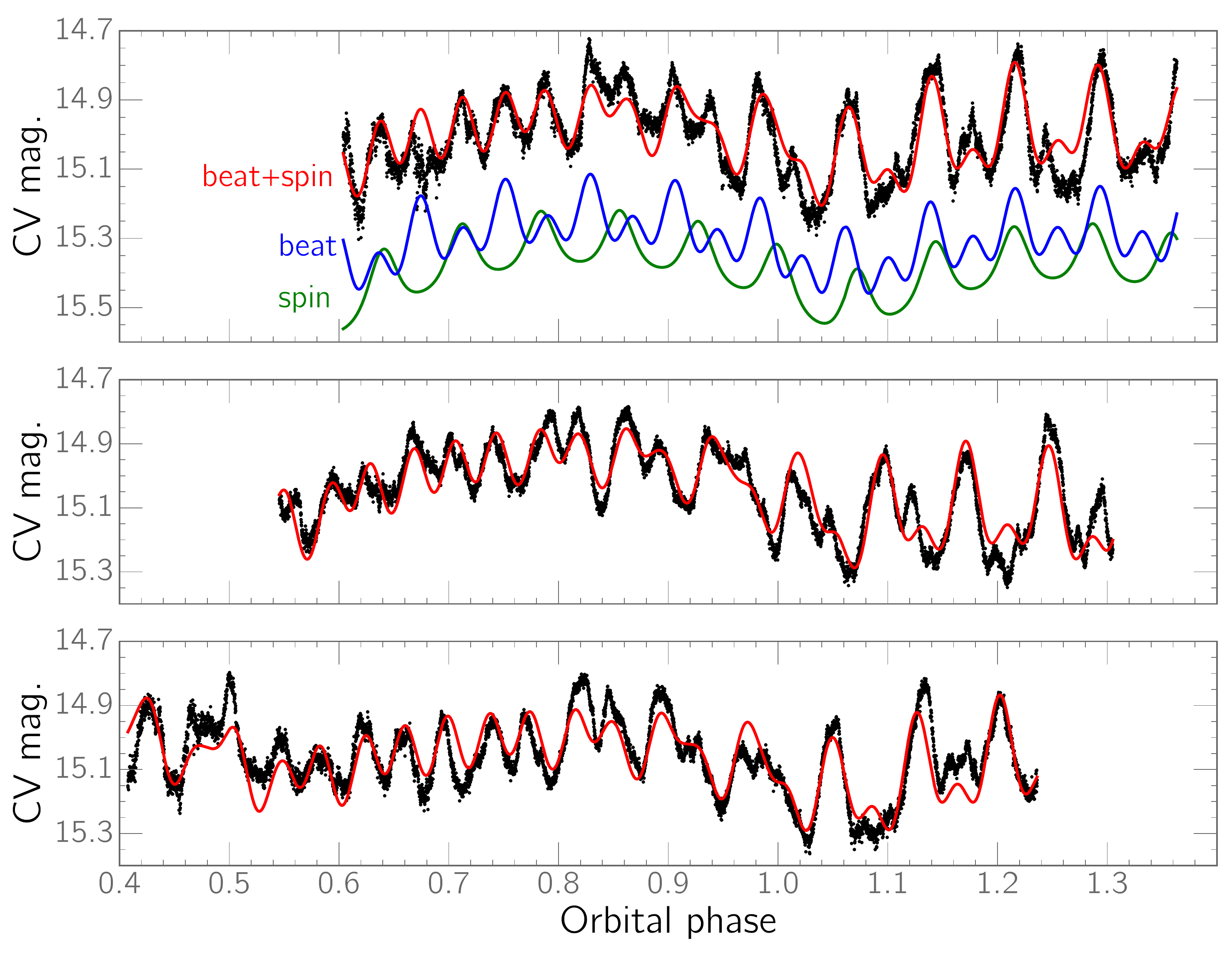}
\caption{The three MDM light curves, obtained on three consecutive nights between JD 2457578-80 with a cadence of 2 sec. The red line in each panel shows our model of the light curve, which is the sum of three components: a beat pulse, a spin pulse, and an orbital variation. For illustration, the top panel plots the beat and spin components of the model in blue and green, respectively, with a constant magnitude offset added to each for clarity. At $\phi_{orb}\sim 0.9$, the frequency of the modulation switches from $2\omega-2\Omega$ to $\omega-\Omega$, a transition explained by the superposition of the beat and spin pulses. Our model predicts that the spin pulse is in phase with one of the beat maxima at orbital phase 0.22 and the other at orbital phase 0.72.  \label{lightcurve}} 

\end{figure*}

FO Aquarii (hereinafter, FO Aqr) is a well-studied IP that has been dubbed the `King of the Intermediate Polars' \citep{patterson83} due to its bright apparent magnitude and large spin pulse amplitude. Over its history, the optical light of FO~Aqr has been dominated by a 20.9-minute spin period, although $\omega -\Omega$, $2\omega-2\Omega$, and various harmonics are also detected, albeit with much less power \citep{kennedy}. Prior to 2016 May, it had been observed exclusively in a high state ($V < 14$). \citet{gs88} found no faint states of FO~Aqr in the Harvard Plate Collection, which contains observations of FO Aqr from 1923-1953, and none are present in long-term data from both the AAVSO and the Catalina Real-Time Sky Survey \citep[CRTS; ][]{drake}. Indeed, there are no published reports of a low state prior to 2016, implying a relatively stable and elevated rate of mass transfer for a long period of time. However, after FO Aqr emerged from solar conjunction in 2016 May, the system was unexpectedly detected near $V\sim15.7$ (see Fig. ~\ref{longterm}), suggesting a major decrease in the mass-transfer rate \citep{atel1}. Thereafter, it gradually rebrightened at an average rate of 0.01 mag day$^{-1}$ between 2016 May and 2016 July and showed a strong 11.26-minute periodicity in its optical light curve \citep{atel2}. Here, we present time-resolved photometry of this archetypal IP's recovery from its unprecedented low state.

\begin{figure*}[ht!]

\epsscale{1.2}
\plotone{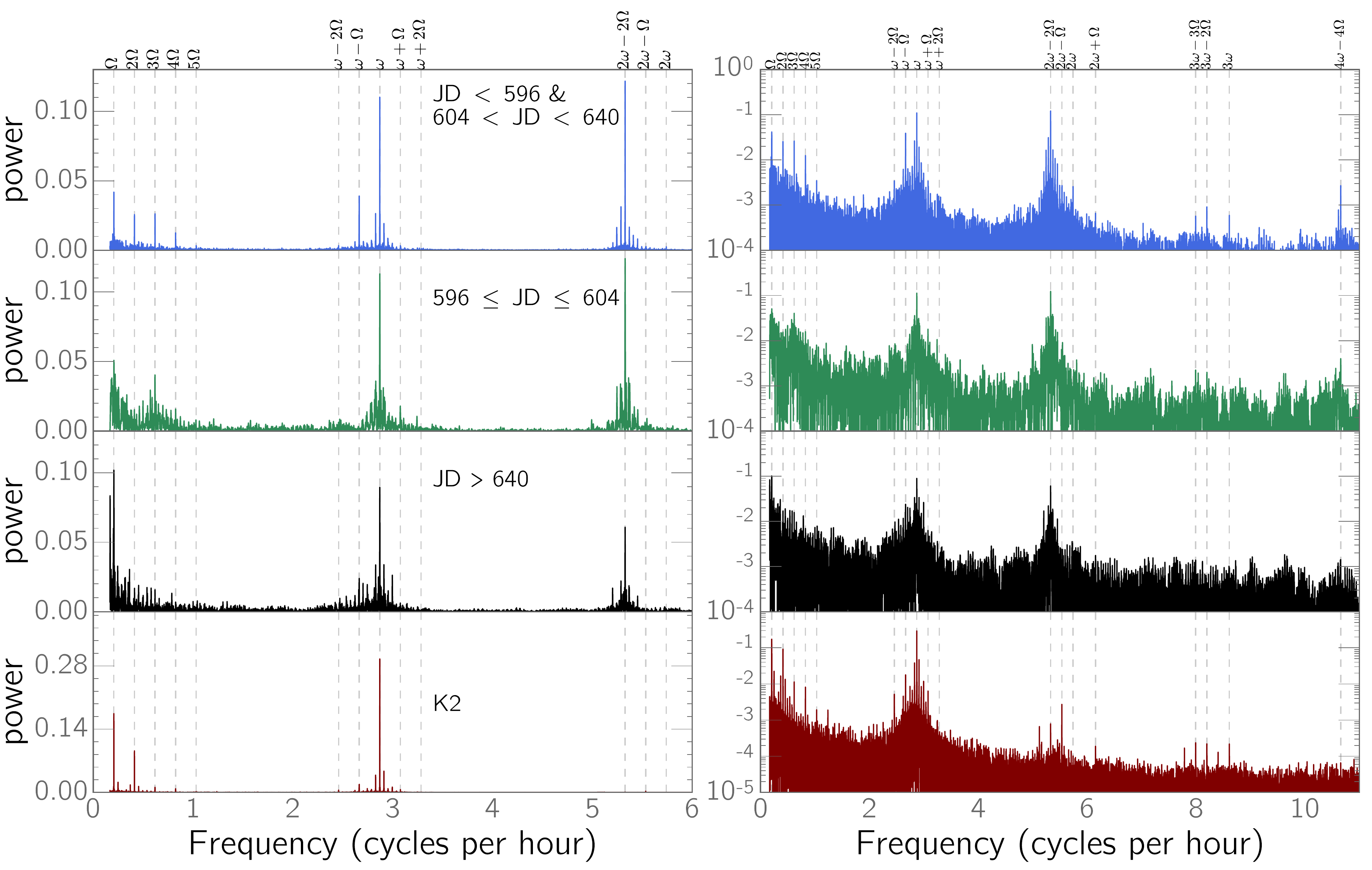}
\caption{ Lomb-Scargle power spectra of FO Aqr during three stages of its recovery (top three rows) compared to the resampled \textit{K2} data from the high state (bottom row). The data are shown with a linear scale in the left column and with a logarithmic scale in the right column. JD refers to JD - 2457000. The top panel shows that for the vast majority of our observations, $2\omega-2\Omega$ was stronger than $\omega$, while $\omega-\Omega$ was clearly detected as well. In comparison to the high state power spectrum (bottom panel), $\omega-\Omega$ and especially $2\omega-2\Omega$ were dramatically stronger during the low state. However, during two segments of the recovery, the power spectrum changed, emphasizing that the recovery is not static. The second row shows an eight-day stretch characterized by a near-total disappearance of $\omega-\Omega$, and the power spectra in the third row feature an increased signal at $\Omega$ along with a diminution of $2\omega-2\Omega$ relative to $\omega$. \label{powerspectra}}

\end{figure*}

\section{Observations}

We obtained unfiltered photometry of FO~Aqr with the 80-cm Sarah L. Krizmanich Telescope (SLKT) at the University of Notre Dame after it emerged from solar conjunction in 2016 May. We also observed FO~Aqr with the Vatican Advanced Technology Telescope (VATT) in 2016 June. These data, together with observations by AAVSO observer Shawn Dvorak made before and after solar conjunction, showed that FO~Aqr had faded by over 2 magnitudes between 2015 December and 2016 May, when we detected it at $V\sim15.7$ \citep{atel1}. Fortuitously, the All-Sky Automated Survey for Supernovae \citep[ASAS-SN; ][]{shappee} observed FO~Aqr twice in 2016 April, and both measurements showed the system over 1 magnitude fainter than normal just two weeks before our detection of the system at $V\sim15.7$. Other than these two ASAS-SN measurements, we are not aware of any other observations of FO Aqr which bridge the gap in coverage between the end of 2015 December and the beginning of 2016 May.

To study FO~Aqr's low-state behavior, we combined unfiltered, time-series photometry from the 1.3-m McGraw-Hill Telescope at MDM Observatory (at a cadence of 2 sec), the 1m Epsilon telescope at the C2PU facility at Observatoire de la C\^{o}te d'Azur, Calern site, France (at a cadence of 20 sec), and the SLKT (at cadences of 6-8 sec). Our dataset also includes all photometric time series with a median cadence of less than 75 sec. from the American Association of Variable Star Observers. All data reported in this paper were obtained before 2016 September 27.

After converting all image timestamps to barycentric Julian date in barycentric dynamical time \citep{eastman}, we performed aperture photometry on the data. Although the data were unfiltered, we determined approximate $V$ magnitudes for FO~Aqr by using $V$ magnitudes of photometric comparison stars when performing photometry.

\section{Analysis}

\subsection{Light Curves \& Power Spectra} \label{LCsec}

Modulations with periods between 11 and 23 minutes are present in all of our light curves, as is a longer-period variation at half of the 4.85-hour orbital period. For most of our observations, the period of the oscillations is strongly dependent on the binary orbital phase ($\phi_{orb}$); see Fig.~\ref{lightcurve}. To compute $\phi_{orb}$, we used the time of orbital minimum from \citet{kennedy} and the orbital period from \citet{md96}. Between $0.9 < \phi_{orb} < 1.4$, the light curve shows a 22.52-min period consisting of two humps of unequal magnitude. During the remainder of the orbital period, the two pulses become approximately equal in amplitude, so the period is halved to 11.26~min. This orbital phase dependence became somewhat less obvious during a bright bump in the recovery between approximately JD 2457596 and 2457604, and the reason for this is discussed in Sec.~\ref{PSsec}.

We computed periodograms of our dataset using both the Lomb-Scargle (\citealt{Lomb76}; \citealt{scargle82}) and phase-dispersion-minimization (PDM; \citealt{Stellingwerf78}) techniques. To prevent high-cadence time series from dominating the power spectra, we resampled each light curve to a cadence of 75 seconds per image. Both Lomb-Scargle and PDM show strong signals at the spin ($\omega$), beat ($\omega - \Omega$), and double-beat ($2\omega - 2\Omega$) frequencies, along with additional sidebands. The Lomb-Scargle method shows the most power at $2\omega - 2\Omega$ (the 11.26 min period), followed by $\omega$, $\Omega$, and $\omega - \Omega$ (the 22.5 min period). The first four harmonics of $\Omega$ are all clearly detected, as are the first two harmonics of $\omega$. The $3\omega - 3\Omega$, $4\omega - 4\Omega$, and $\omega+\Omega$ frequencies also show some power. Although the PDM technique also favors $2\omega - 2\Omega$ as the strongest signal, it shows $\omega - \Omega$ to be stronger than $\omega$. The Lomb-Scargle periodogram of the low state is shown in the top panel of Fig.~\ref{powerspectra}, along with a power spectrum of high-state data for comparison. 

The power spectrum varied during the recovery, as shown in the middle two rows of Fig.~\ref{powerspectra}. During the bump in the long-term light curve between JD 2457596 and 2457604, the power at $\omega-\Omega$ was almost negligible. Additionally, after JD 2457604, the power at $\Omega$ increased significantly, as did $\omega$ in comparison to $2\omega-2\Omega$.

\textit{Kepler K2} observations of FO~Aqr's high state \citep{kennedy} provide a useful benchmark for comparison. Since the \textit{K2} data were obtained at a one-minute cadence for 69 consecutive days, we resampled the \textit{K2} data using the following algorithm in order to approximate our own sampling rate. The \textit{K2} and low-state data were each stored in a separate array. For each observation in the \textit{K2} and low-state datasets, we calculated the time elapsed since the earliest observation in that dataset. Then, for each elapsed-time value in the low-state data, we found the closest elapsed-time value in the \textit{K2} data. The timestamp and magnitude of this \textit{K2} measurement were then included in the resampled light curve. The algorithm prevented an observation from being added to the resampled light curve more than once.

There are a number of striking differences between our power spectra and the resampled \textit{K2} power spectra, underscoring that the accretion processes have fundamentally changed in response to the diminished mass-transfer rate. In the resampled high-state data, $\omega$ was by far the strongest signal, followed by $\Omega$. The \textit{K2} data also show $\omega - \Omega$ and especially $2\omega - 2\Omega$ to be weak in comparison to $\omega$. The signal at $2\omega - 2\Omega$ shows the most dramatic difference between the low and high states, increasing in power by well over an order of magnitude to become the dominant signal of the low-state power spectrum. In short, FO Aqr has transitioned from an $\omega$-dominated power spectrum in its high state to one dominated by $2\omega-2\Omega$ in the low state.

\subsection{Model Light Curve}

The power spectra are helpful in identifying which frequencies are present in the data, but they do not provide information on phases of the components. Thus, we model the light curves based on the periodicities observed in the power spectra.

To approximate the beat ($\omega-\Omega$) and double-beat ($2\omega-2\Omega$) signals that dominate the low-state light curves, we add two periodic functions, one shifted in phase by $\pi$ relative to the other. Because we do not know the true shape of the pulses, we have tested several possible periodic functions including
$F=(\cos(\theta)+1)^2/4$ and
$F=\cos(\theta)^3$, but the choice makes little difference in matching the observations. We also include a spin pulse and an orbital modulation. The full model is:

\begin{multline*}
\label{modelspin}
\begin{split}
F_{tot} &= A_{s} F_{s}\left(t,\omega,\phi_{s} \right) + A_{b1} F_{b}\left(t,\omega-\Omega,\phi_{b} \right)\\
  &+ A_{b2} F_{b}\left(t,\omega-\Omega,\phi_{b}+\pi \right) \\
  &+ A_{o} \left( F_{o}\left(t,\Omega,\phi_{o} \right) + A_{o2} F_{o}\left(t,\Omega,\phi_{o} + \pi \right) \right)\; ,
\end{split}
\end{multline*}
where $F_{s},F_{b}$ and $F_{o}$ are all periodic functions describing the spin pulse, beat pulse, and orbital modulations seen in the light curve. When fitting the light curves, we fix the orbital and spin periods at 0.20205976~d and 1254.3401~s respectively. We then calculate
the least-squares difference between the data and the model by varying the component amplitudes ($A_s$, $A_{b1}$, $A_{b2}$, $A_o$) and the phases ($\phi_s$, $\phi_b$, $\phi_o$). $A_{o2}$ allows freedom in fitting the amplitudes of the 2$\Omega$ component, and appears close to unity in our light curves.

Because of their very high signal-to-noise ratio, we used the MDM data to estimate the model parameters, and the resulting fit is shown in Fig.~\ref{lightcurve}. The model works quite well in matching the brightness and times of the peaks. The ratio between the model amplitudes ($A_s$:$A_{b1}$:$A_{b2}$) is
1.0:2.4:1.5. While the components are roughly comparable in amplitude, the spin varies in phase with respect to the double-humped beat signal. In the first half of the orbit, the spin pulse is nearly in phase with the stronger beat maximum, making the $\omega-\Omega$ variation stand out. In the second half of the orbit, the spin pulse occurs at a similar phase to the lower amplitude beat maximum, creating a prominent $2\omega-2\Omega$ variation. Because of the relation between the orbital, spin, and beat periods, this pattern repeats each orbit. Our modeling shows that the spin pulse is approximately in phase with the stronger beat maximum at $\phi_{orb} \sim 0.22$ and with the weaker one at $\phi_{orb} \sim 0.72$.

\begin{figure}

\epsscale{1.2}
\plotone{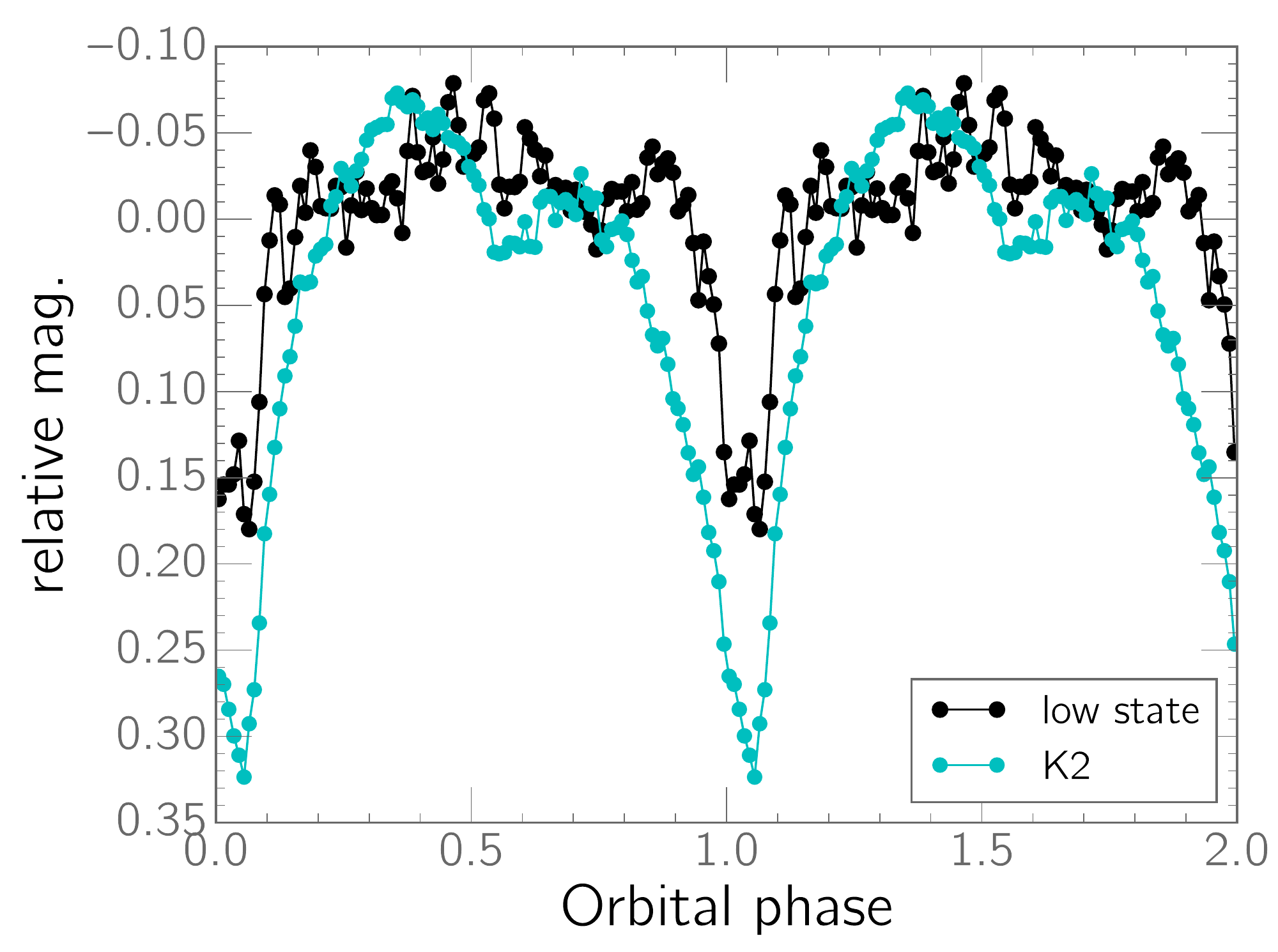}
\caption{A phased light curve of the orbital period in the low state. The bins are non-overlapping and are 0.01 phase units wide. To show the differences compared to the high state, the figure also shows an orbital phase plot using \textit{K2} observations.\label{phaseplot}}

\end{figure}

\section{Discussion}

\subsection{Long-Term Light Curve}

The drop in brightness is likely a result of a major decrease in the mass transfer rate in the system. This can cause a decrease in the disk luminosity as well as the brightness of the accretion curtains and the shock energy at the WD accretion poles. \citet{livio94} suggested that the cause of low states in magnetic CVs could be starspots on the secondary passing over the inner Lagrangian point. The paucity of low states in the long-term light curve of FO~Aqr implies that magnetic activity on its secondary is weaker than in other magnetic CVs. 

The rate at which a CV fades to a low state or brightens to a high state is often quantified by its $e$-folding time, which is the time required for the system's flux to change by a factor of $e$. \citet{kafka05} studied 43 low-states in five magnetic CVs and found that the typical transition $e$-folding time-scale was 20~days, or a rate of -0.05 mag~day$^{-1}$. This is significantly faster than the measured $e$-folding time for FO~Aqr's recovery (115$\pm7$~days), which we found using an outlier-resistant linear regression of the low-state data shown in Fig.~\ref{longterm}. Indeed, the $e$-folding time for FO~Aqr's recovery is over twice as long as the slowest single-sloped transition from \citet{kafka05}. Although some transitions occur in two steps with different recovery rates, the slowest $e$-folding time reported by \citet{kafka05} during the faintest part of a recovery is just 66 days. We conclude, therefore, that the recovery rate from the low state of FO~Aqr has been unusually slow.

The two ASAS-SN observations of FO~Aqr in an intermediate state in 2016 April, in conjunction with the earliest ASAS-SN observation of the low state in 2016 May, enable us to constrain the upper limit of the decline's $e$-folding time. We assume that the two measurements from April captured the original decline from the high state and that the decline was smooth and single-sloped. Since these data are contaminated by the variability at $\omega$, $\Omega$, and the associated sidebands, we used Monte Carlo simulations to simulate the effect of these modulations on the decline's $e$-folding time. We find that for a single-sloped transition, the $e$-folding time must have been less than $17^{+5}_{-3}$ days. Since the rebrightening was underway just two weeks after the two April observations (Fig.~\ref{longterm}), recovery from the low state must have started almost immediately after the end of the initial fade to minimum brightness.

While starspots at the L1 point may explain the decrease in the mass transfer rate and the concomitant fading of the disk and accretion luminosity, the reason for the slow recovery is not clear. To explain the VY~Scl class of CVs, \citet{wu95} theorized that loss of shielding by an accretion disk can heat the secondary, thus raising the mass transfer rate. So for FO~Aqr, we expect that the partial depletion of the disk would result in extra heating of the secondary, leading to an acceleration of the recovery. \citet{wu95} expect mass transfer and disk building should occur on a timescale of 10$^6$~s. If the mass transfer through the L1 point had recovered to its pre-2016 rate, we would have expected the luminosity rebound to be on the viscous timescale of the disk. The very slow recovery rate we observe suggests that the secondary star has yet to return to its previous mass-loss rate.

As is evident from the bottom panel of Fig.~\ref{longterm}, the rate of brightening during the recovery was not perfectly uniform. For example, between JD~2457596 and JD~2457604, FO~Aqr deviated above the long-term trend by about 0.2~mag. During this ``bump," flares reaching magnitude 14.3 were observed in the time-resolved photometry, suggesting irregular accretion events.

\subsection{Orbital Modulation} \label{Orbitsec}

In the bright state, FO~Aqr shows a dip in brightness every 4.85~hr which is thought to be a grazing eclipse of a disk \citep{cmc89}. We constructed a phased light curve of our entire dataset to determine if this feature is also present in the low state. To prevent the high-cadence data from dominating this plot, we resampled each light curve into 75-second bins. We then split the resampled data into 100 non-overlapping orbital-phase bins and calculated the median magnitude in each of those bins. Although the signals at $\omega$, $\omega-\Omega$, and $2\omega-2\Omega$ usually overwhelm the orbital modulation in individual light curves, they are all out of phase at the orbital period. Thus, the inclusion of data from many different orbital cycles smooths out these variations, isolating the comparatively weak orbital signal. 

The resulting phase plot (Fig.~\ref{phaseplot}) shows an eclipse near the orbital phase expected from the high-state light curve \citep{kennedy}, strongly suggesting that there was still an accretion disk during our observations. The low-state eclipse depth decreased by $\sim$0.15 mag compared to the \textit{K2} data, as would be expected if the disk were depleted during the low state. The eclipse width is also significantly less in the low state (FWHM = $0.104\pm0.005\phi_{orb}$) than in the high state (FWHM = $0.18\pm0.007\phi_{orb}$), implying that the outer radius of the disk has shrunk (K. Mukai, private communication).

\subsection{Spin and Beat Signals}
\label{PSsec}

Power spectral analysis of an IP can serve as an important method for determining whether the system is undergoing disk-fed accretion, stream-fed accretion, or some combination of the two \citep{fw99}. Fig.~\ref{powerspectra} clearly shows that the double-beat ($2\omega - 2\Omega$) frequency dominates the optical light curve during the low state of FO~Aqr, with $\omega - \Omega$ also being significantly stronger than in the high state. Moreover, the model fit to the MDM light curve shows that the beat amplitude is more than twice the strength of the spin component, while in the high state the beat amplitude is only a fifth of the spin amplitude. In accordance with the modeling in \citet{fw99}, the increased strength of the signals at the beat and double beat suggests that a substantial fraction of the accretion during our observations was stream-fed.

Although $\omega$ is quite strong in our power spectra, \citet{fw99} predict that $\omega$ is not uniquely indicative of disk-fed accretion. They caution that stream-fed accretion can induce a signal at $\omega$ depending on several effects, including the optical depth of the curtains, the extent of the accretion stream, and the strength of cyclotron radiation from the accretion region. Thus, the presence of $\omega$ in the low-state power spectrum does not conclusively establish that there was simultaneous disk-fed accretion. Low-state X-ray and spectral observations should reveal more definitively whether the WD was accreting significantly from the disk (Kennedy et al., in prep.).

An alternate explanation for the signals at $\omega - \Omega$ and $2\omega - 2\Omega$ is that they result from the irradiation of stationary structures within the binary rest frame ({\it e.g.}, a disk bulge or the donor star). High-energy photons from one of the WD's accretion regions will sweep over these structures at a frequency of $\omega - \Omega$. In FO~Aqr's high state, this mechanism causes a modulation at that frequency \citep{osborne}. A difficulty with this scenario is that during the low state, the disk would have been significantly depleted by the low mass-transfer rate, as we infer from the decreased eclipse depth and width during the low state (Sec.~\ref{Orbitsec}). While we favor stream-fed accretion as the cause of the $\omega-\Omega$ and $2\omega-2\Omega$ variations, their origin can likely be established with greater confidence from time-resolved spectroscopy of the beat modulation (Kennedy et al., in prep.).

During the ``bump" in the long-term recovery of FO~Aqr, the power of the $\omega - \Omega$ signal weakened, as shown in Figure~\ref{powerspectra}. Outside of the bump, the two maxima of the beat modulation were asymmetric, leading to a significant signal at the beat frequency. The strength of the $\omega - \Omega$ signal declined during the bump simply because the amplitudes of the beat pulses became comparable to each other, causing the power to move to the 2$\omega - 2\Omega$ frequency. We expect that the beat pulses result from accretion onto both magnetic poles, with the relative brightness of each beat maximum determined by the accretion rate through the accretion curtains and by the amount of obscuration by the disk. For example, at FO~Aqr's normal brightness, the $2\omega - 2\Omega$ signal is greatly suppressed, possibly because the fully engorged disk blocks our view of one pole.

\begin{figure}

\epsscale{1.22}
\plotone{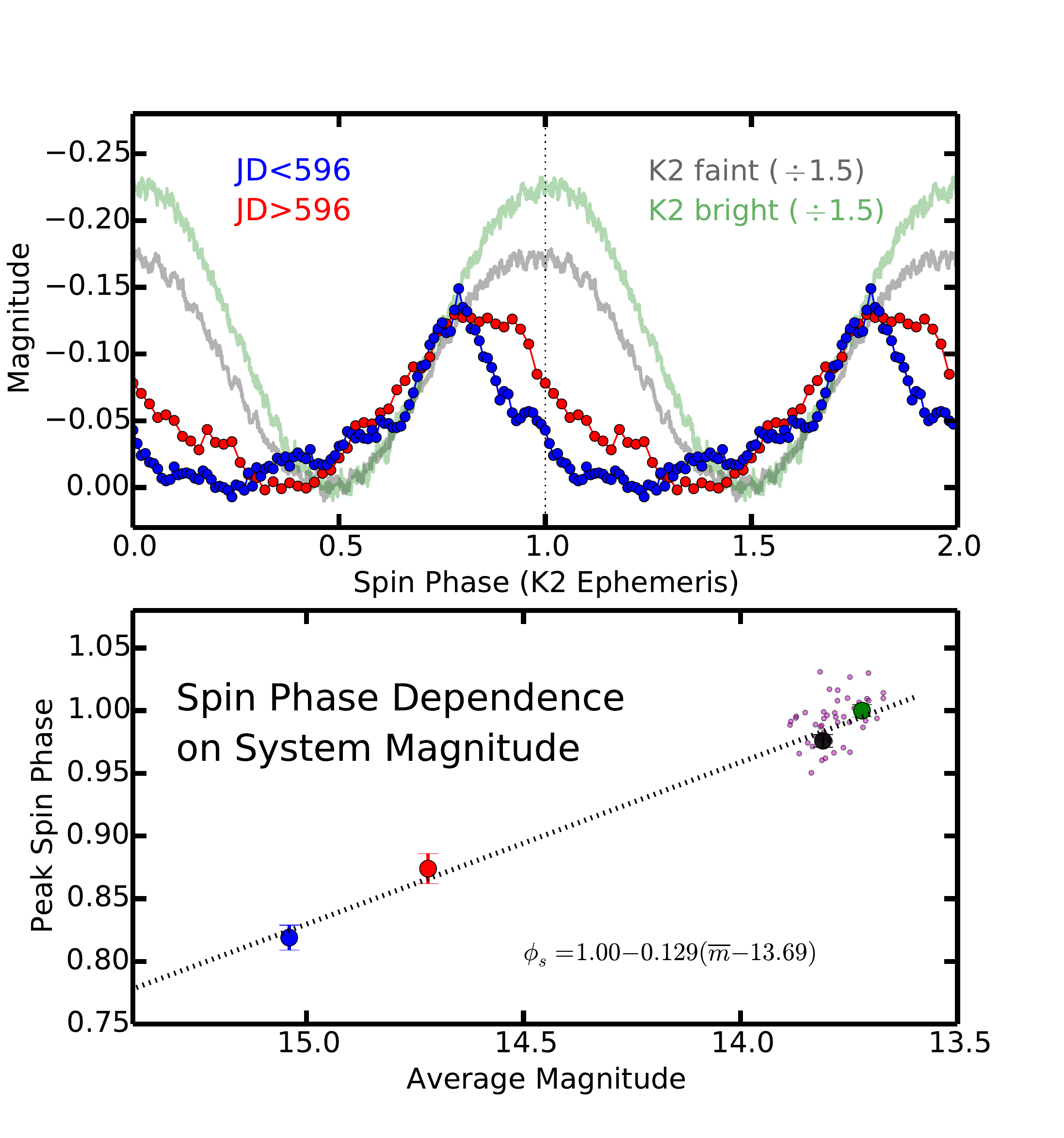}
\caption{{\bf Top:} Light curves phased on the spin period and then smoothed with a median filter. Phased light curves obtained before JD~2457596 are shown in blue and those obtained between 2457596 and 2457604 are shown in red. The spin-phased light curve based on the brightest third of the \textit{K2} observations is shown in green while the faintest third is plotted in gray. The time of the spin peak is clearly shifted in the low state relative to the \textit{K2} data. {\bf Bottom:} The phase of the spin peak plotted against the median magnitude, showing an approximately linear relationship. The small dots show the spin phase and FO~Aqr brightness calculated by dividing the \textit{K2} observations into fifty subsets. As described in the text, these figures have been corrected for a small phase shift caused by a change in the spin period. \label{spinphase}}

\end{figure}

\subsection{Orbital-Phase Dependence of Spin Pulses}

\citet{kennedy} noted that the phase of the spin pulse in the \textit{K2} data was mildly dependent on the brightness of FO~Aqr. If we assume that $\omega$ is the pulse from a disk-fed accretion curtain,\footnote{See Sec.~\ref{PSsec} for an alternative interpretation of $\omega$.} then our analysis shows that the spin-pulse arrival times are indeed dependent on the system's overall brightness. We isolated the spin pulse in our photometry by calculating the spin phase for each observation using the ephemeris 
$$T_{spin}[BJD] = 2457032.54909(1)+0.014517825(5)\times E, $$
based on the \citet{kennedy} spin period. We sorted the photometry by spin phase before smoothing the phased light curve with a median filter. The resulting spin light curves are shown in Fig.~\ref{spinphase}. 

By using this technique to measure the phase of the spin peak in the \textit{K2} data, we find that the spin pulse arrived $-0.024\pm 0.007$ earlier in phase during the faintest third of the \textit{K2} data than it did during the brightest third (a difference of 0.11 mag). In the low-state photometry, our light curves taken before JD~2457596 (when FO~Aqr had an average unfiltered magnitude of 15.0) show that the spin pulse arrived 4.3~minutes early ($-0.20\pm 0.01$ in phase). This is significantly larger than the expected K2 ephemeris uncertainty of 17~s. Between JD~2457596 and JD~2457604 (when FO~Aqr had brightened to an average of 14.7~mag), the spin peak arrived 3.1~minutes ($-0.15\pm 0.01$ in phase) earlier than the \textit{K2} ephemeris. To test whether a change in the spin period can account for these shifts, we constructed a cubic ephemeris for the spin maxima using recent data in Table~1 in \citet{B16}, and we compute a total phase shift of $-0.024$ phase units between the \textit{K2} observations and the low state. This accounts for only a small fraction of the observed phase shift of $-0.20\pm 0.01$ during that span, so it is unlikely that the phase shift is the result of a change in the spin period of the WD. Indeed, we can connect our data and the \textit{K2} with an approximately linear relation between the average magnitude of the system and the phase of the spin pulse (Fig.~\ref{spinphase}, bottom panel).

\citet{hmb97} provide a possible explanation for this dependence. They found that during an outburst of the intermediate polar XY~Ari, the inner radius of the disk decreased by over $\sim$50\% because of the increased rate of mass transfer ($\dot{M}$) through the inner disk. Thus, they concluded that the gas in the disk was captured by different field lines than during quiescence, resulting in a pronounced shift in both the location and size of the accretion region. This scenario is directly applicable to FO~Aqr. In response to the decreased $\dot{M}$ during FO~Aqr's low state, there might have been an increase in the radius at which the disk is magnetically truncated, thereby altering the shape and size of the disk-fed accretion curtains. This effect could also potentially explain why the spin pulse in FO~Aqr appears to become wider and more sinusoidal as the system brightens. X-ray light curves of the low state provide one possible mechanism for testing whether the accretion region has changed in comparison to the high state.

The phase shift of the spin pulses has significant implications for previous studies of FO~Aqr's spin period. For decades, the spin timing of FO~Aqr has been used to identify changes in the WD spin period \citep[e.g.,][]{patterson94}. However, the spin pulse's dependence on system brightness might require a correction to be made in order to prevent the phase shift from masquerading as a change in the spin period. As Fig.~\ref{spinphase} shows, this effect is apparent even in the relatively minor brightness variations in the high-state \textit{K2} data. Consequently, it will be important to monitor the phase of the spin pulse as the recovery continues so that this dependence may be tested more rigorously.

\section{Conclusion}

We show that FO~Aqr was nearly 2.5~mag fainter than normal and at its lowest recorded brightness when it emerged from solar conjunction in 2016 May. The light curve in its low state was significantly different from the spin-dominated variability observed when the system is bright. While photometric variations at the spin frequency are still present in the low state, the dominant signal is at twice the beat frequency. This suggests that FO~Aqr was accreting substantially from its accretion stream during the system's low state. In addition, the $e$-folding time of the recovery ($115\pm7$ days) is unusually slow. The observed recovery is longer than the viscous timescale of the disk, implying that the mass-transfer rate from the secondary has not yet returned to normal. The decreased depth and width of the disk-grazing eclipse suggest that the disk has been at least partially depleted. Finally, we find that the arrival times of the spin pulse are dependent upon the system's overall brightness. It will be important for future studies to determine whether the pulse at $\omega$ is indeed attributable to disk-fed accretion curtains or is instead from other processes that can produce a signal at $\omega$ \citep{fw99}. Nevertheless, the presence of the grazing eclipse during the low state shows that there was at least one prerequisite for disk-fed accretion---namely, a disk.

\acknowledgments

We appreciate the expeditious review and helpful comments by the anonymous referee. We are grateful to Koji Mukai for his very thoughtful comments on an earlier version of this manuscript.

The Sarah L. Krizmanich Telescope was a generous donation to the University of Notre Dame in memory of its namesake by the Krizmanich family.

This work is based on observations obtained at the MDM Observatory, operated by Dartmouth College, Columbia University, Ohio State University, Ohio University, and the University of Michigan.

We thank the Vatican Observatory for granting observing time on the Vatican Advanced Technology Telescope.

This work was partially supported by the National Science Foundation under award AST-1411685 to the Ohio State University.

BS is supported by NASA through Hubble Fellowship grant HF-51348.001 awarded by the Space Telescope Science Institute, which is operated by the Association of Universities for Research in Astronomy, Inc., for NASA, under contract NAS 5-26555. 

We thank LCOGT and its staff for their continued support of ASAS-SN.  Development of ASAS-SN has been supported by NSF grant AST-0908816 and CCAPP at the Ohio State University. ASAS-SN is supported by NSF grant AST- 1515927, the Center for Cosmology and AstroParticle Physics (CCAPP) at OSU, the Mt. Cuba Astronomical Foundation, George Skestos, and the Robert Martin Ayers Sciences Fund.

PG, MRK and PC acknowledge support from the Naughton Foundation and the University College Cork Strategic Research Fund.

DC acknowledges the assistance of Pham Nguyen and Max Morehead.

This work has utilized photometry from the AAVSO International Database, contributed by observers worldwide. We thank the observers who have contributed observations of FO Aqr, including Valery Tsehmeystrenko, Roger Pickard, Luis Perez, Nathan Krumm, and Lewis Cook. We also thank Elizabeth Waagen for her assistance with organizing an AAVSO campaign on FO~Aqr.


\begin{thebibliography}{}


\bibitem[Bonnardeau(2016)]{B16} Bonnardeau M. 2016, IBVS, 6181, 1

\bibitem[Drake et al.(2009)]{drake} Drake, A.~J., Djorgovski, S.~G., Mahabal, A., et al. 2009, \apj, 696, 870

\bibitem[Eastman, Siverd, \& Gaudi(2010)]{eastman} Eastman, J., Siverd, R., Gaudi, B.~S. 2010, PASP, 122, 935 

\bibitem[Ferrario \& Wickramasinghe(1999)]{fw99} Ferrario, L. \& Wickramasinghe, D.~T. 1999, MNRAS, 309, 517

\bibitem[Garnavich \& Szkody(1988)]{gs88} Garnavich, P. \& Szkody, P. 1988, PASP, 100, 1522

\bibitem[Hameury et al.(1986)]{hameury86} Hameury J. M., King A. R. \& Lasota J. P., 1986, MNRAS, 218, 695

\bibitem[Hellier, Mukai, \& Beardmore(1997)]{hmb97} Hellier, C., Mukai, K., \& Beardmore, A.~P. 1997, MNRAS, 292, 397

\bibitem[Hellier, Mason, \& Cropper(1989)]{cmc89} Hellier, C., Mason, K.~O., \& Cropper, M. 1989, MNRAS, 237, 39

\bibitem[Kafka \& Honeycutt(2005)]{kafka05} Kafka, S., \& Honeycutt, R.~K.\ 2005, \aj, 130, 742 

\bibitem[Kennedy et al.(2016)]{kennedy} Kennedy, M. R., Garnavich, P. Breedt, E., Marsh, T. R., Gansicke, B. T., Steeghs, D., Szkody, P., \& Dai, Z. 2016, MNRAS, 459, 3622

\bibitem[Littlefield et al.(2016a)]{atel1} Littlefield, C., Garnavich, P., Aadland, E., \& Kennedy, M. 2016a, ATel, 9216, 1

\bibitem[Littlefield et al.(2016b)]{atel2} Littlefield, C., Aadland, E., Garnavich, P. \& Kennedy, M. 2016b, ATel, 9225, 1

\bibitem[Livio \& Pringle(1994)]{livio94} Livio, M., \& Pringle, J. E. 1994, ApJ, 427, 956

\bibitem[Lomb(1976)]{Lomb76} Lomb, N.~R. 1976, Astrophysics \& Space Science, 39, 447-462

\bibitem[Osborne \& Mukai(1989)]{osborne} Osborne, J.~P. \& Mukai, K. 1989, MNRAS, 238, 1233

\bibitem[Marsh \& Duck(1996)]{md96} Marsh, T.~R. \& Duck, S.~R. 1996, NewA, 1, 97

\bibitem[Patterson \& Steiner(1983)]{patterson83} Patterson J. \& Steiner J. E., 1983, ApJ, 264, L61

\bibitem[Patterson(1994)]{patterson94} Patterson, J. 1994, PASP, 106, 209

\bibitem[Rosen et al.(1988)]{rosen88} Rosen S. R., Mason K. O., C\'{o}rdova F. A. 1988, MNRAS, 231, 549

\bibitem[Scargle(1982)]{scargle82} Scargle, J.~D. 1982, ApJ, 263, 835

\bibitem[Shappee et al.(2014)]{shappee} Shappee, B. et al. 2014, \apj, 788, 48

\bibitem[Stellingwerf(1978)]{Stellingwerf78}Stellingwerf, R. 1978, ApJ, 224, 953

\bibitem[Warner(1986)]{warner86} Warner, B. 1986, MNRAS, 219, 347

\bibitem[Wu et al.(1995)]{wu95} Wu, K., Wickramasinghe, D.~T., \& Warner, B.\ 1995, \pasa, 12, 60 

\end{thebibliography}
\end{document}